# Self-Confirming Price Prediction for Bidding in Simultaneous Ascending Auctions

Anna Osepayshvili, Michael P. Wellman, Daniel M. Reeves, and Jeffrey K. MacKie-Mason
University of Michigan
Ann Arbor, MI 48109 USA
{annaose, wellman, dreeves, jmm}@umich.edu

## Abstract

Simultaneous ascending auctions present agents with the *exposure problem*: bidding to acquire a bundle risks the possibility of obtaining an undesired subset of the goods. Auction theory provides little guidance for dealing with this problem. We present a new family of decision-theoretic bidding strategies that use probabilistic predictions of final prices. We focus on *self-confirming price distribution* predictions, which by definition turn out to be correct when all agents bid decision-theoretically based on them. Bidding based on these is provably not optimal in general, but our experimental evidence indicates the strategy can be quite effective compared to other known methods.

## 1 Simultaneous Ascending Auctions

A *simultaneous ascending auction* (SAA) (Cramton, 2005) allocates a set of $M$ related goods among $N$ agents via separate, concurrent English auctions for each good. Each auction may undergo multiple rounds of bidding. At any given time, the *bid price* on good $m$ is $\beta_m$, defined to be the highest non-repudiable bid $b^m$ received thus far, or zero if there have been no bids. To be admissible, a new bid must meet the *ask price*, i.e., the bid price plus a bid increment (which we take to be one w.l.o.g.), $b^m_{new} \geq \beta_m + 1$. If an auction receives multiple admissible bids in a given round, it admits the highest (breaking ties arbitrarily). An auction is *quiescent* when a round passes with no new admissible bids. Upon mutual quiescence, the auctions close and allocate their respective goods per the last admitted bids.

Because no good is committed until all are, an agent's bidding strategy in one auction cannot be contingent on the outcome for another. Thus, an active agent desiring a bundle of goods runs the risk that it will purchase some but not all goods in the bundle. This is known as the *exposure problem*, and arises whenever agents have complementarities among goods allocated through separate markets.

One approach to exposure is to design mechanisms that take the complementarities directly into account, such as *combinatorial auctions* (Cramton et al., 2005; de Vries and Vohra, 2003), in which the auction mechanism determines optimal packages based on agent bids over bundles. Although such mechanisms may provide an effective solution in some cases, there are often significant barriers to their application (MacKie-Mason and Wellman, 2005).

We design bidding strategies to perform well despite exposure risk. Let $v(X)$ denote the value to a particular agent of obtaining the set of goods $X$. Given that it obtains $X$ at prices $\boldsymbol{p}$, the agent's *surplus* is its value less the amount paid, $\sigma(X, \boldsymbol{p}) \equiv v(X) - \sum_{m \in X} p_m$. The agent selects the bundle that maximizes its surplus evaluated at its *perceived prices*, $\hat{\boldsymbol{p}}$:

$$X^* = \arg\max_X \sigma(X, \hat{\boldsymbol{p}}). \qquad (1)$$

Each strategy we analyze is defined by how the agent constructs $\hat{\boldsymbol{p}}$ from its information state. Given that decision, the agent bids $b^m = \beta_m + 1$ (the ask price) for the $m \in X^*$ that it is not already winning.

One example is the widely-studied *straightforward bidding* (SB) strategy.[1] An SB agent uses *myopically perceived prices*: the bid price for goods it was winning in the previous round and the ask price for the others.

An agent has *single-unit preference* iff for all $X$, $v(X) = \max_{m \in X} v(\{m\})$. For such agents, SB is a *no regret* policy (Hart and Mas-Colell, 2000), as the agent would not wish to change its bid even after observing what the other agents did (Bikhchandani and Mamer, 1997). When all agents have single-unit preference, and value every good equally, the situation is equivalent to a problem in which all buyers have an inelastic demand for a single unit of a homogeneous commodity. For this problem, Peters and

---

[1] We adopt the terminology introduced by Milgrom (2000). The same concept is also referred to as "myopic best response", "myopically optimal", and "myoptimal" (Kephart et al., 1998).

Severinov (2001) showed that straightforward bidding is a perfect Bayesian equilibrium. Up to a discretization error, the allocations from SAAs are efficient when agents follow straightforward bidding. It can also be shown (Bertsekas, 1992; Wellman et al., 2001) that the final price vector will differ from the minimum unique equilibrium price by at most $\kappa \equiv \min(M, N)$. The value of the allocation, defined to be the sum of the bidder surpluses, will differ from the optimal by at most $\kappa(1+\kappa)$. Unfortunately, these nice properties do not generalize to other preferences. The final SAA prices can differ from the minimum equilibrium price vector, and the allocation value can differ from the optimal, by arbitrarily large amounts (Wellman et al., 2001). SB's obliviousness to exposure can cause an agent to incur significant losses in cases where these may have been anticipated and avoided.

Despite the fact that markets for interdependent goods operating simultaneously and independently are ubiquitous, auction theory to date (Krishna, 2002) has little to say about how one *should* bid in simultaneous markets with complementarities. SAA-based auctions are even deliberately adopted, despite awareness of strategic complications (Milgrom, 2000), for some markets that are expressly designed, most famously the US FCC spectrum auctions starting in the mid-1990s (McAfee and McMillan, 1996). Simulation studies of scenarios based on the FCC auctions shed light on some strategic issues (Csirik et al., 2001), as have accounts of some of the strategists involved (Cramton, 1995; Weber, 1997), but the general game is still too complex to admit definitive strategic recommendations.

## 2 Background Research

We previously introduced two SB extensions to mitigate the exposure problem in a market for scheduling resources (MacKie-Mason et al., 2004; Reeves et al., 2005). First, we modified SB to approximately account for sunk costs, recognizing that goods an agent is already winning impose no incremental costs if other agents do not submit additional bids (Reeves et al., 2005). We solved for settings of a "sunk awareness" strategy parameter such that agents playing pure or mixed forms of this strategy are in Nash equilibrium. We identified qualitatively distinct equilibrium settings of this parameter corresponding to different preference distributions.

Second, we allowed agents to select bundles based on predicted closing prices for each good (MacKie-Mason et al., 2004). Performance depends on the specific price prediction, so we defined strategies based on various *methods* for predicting. We found that this approach is quite effective compared to the strategies based on a sunk-awareness parameter, including SB.

We now extend point prediction-based strategies to employ *probability distributions*, and extend our notion of self-confirming (SC) prediction to the case of probability distributions. We also substantially improve our empirical game analysis methodology, by developing techniques to embrace large strategy sets without requiring exhaustive examination of the full combinatorial set of strategy profiles. We then use our methodology to analyze a broad range of strategies in multiple environments. The SC distribution-based strategy—though not always optimal—performs very well overall, and we argue it is likely to be difficult to improve upon for general classes of SAAs.

## 3 Probabilistic Price Predictions

When an agent does not have single-unit preference, and chooses to bid on a bundle of size $> 1$, it may face exposure. If at any point the agent is bidding for a bundle $X$ at price such that $\sum_{m \in Y} b^m > v(Y)$ for some $Y \subset X$, the agent is exposed. Exposure in an SAA is a direct tradeoff: bidding on a needed good increases the prospects for completing a bundle, but also increases the expected loss in case the full set of required goods cannot be acquired. A decision-theoretic approach would account for these expected costs and benefits, choosing to bid when the benefits prevail, and cutting losses in the alternative.

Consider the $M = N = 3$ example presented in Table 3. Agents 2 and 3 have single-unit preference. Agent 1 does not, and indeed needs all three goods to obtain any value.

Table 1: An example of agent preferences.

| Name | $v(\{1\})$ | $v(\{2\})$ | $v(\{3\})$ | $v(\{1,2,3\})$ |
|---|---|---|---|---|
| Agent 1 | 0 | 0 | 0 | 15 |
| Agent 2 | 8 | 6 | 5 | 8 |
| Agent 3 | 10 | 8 | 6 | 10 |

If all three agents bid straightforwardly (SB), a possible outcome is that agent 3 wins the first good at 7, agent 1 wins the second at 5, and agent 2 wins the third at 4. Here, agent 1 is caught by the exposure problem, stuck with a useless good and a surplus of $-5$.[2] Agent 1 cannot do better against SB bidders by continuing to bid on the first two slots in this example. In general there is no known Bayes-Nash optimal strategy for SAAs when agent preferences exhibit complementarities, as do agent 1's in Table 3. Thus we focus on analyzing the performance of promising but provably non-optimal strategies.

The effectiveness of a particular strategy will in general be highly dependent on the characteristics of other agents in the environment. Thus, we turn to strategies that employ preference distribution beliefs to guide bidding behavior, rather than relying only on current price information as in the SB strategy.

---
[2]Depending on the sequence of bidding (when asynchronous), and the outcome of random tie-breaking (when synchronous) several different outcomes are possible, with agents following SB. If agent 1 bids at all, however, it ultimately ends up exposed, with negative surplus.

One natural use for preference distribution beliefs is to form *price predictions* for the goods in an SAA. In the example above, suppose agent 1 could predict with certainty before the auctions start that the prices would total at least 16. Then it could conclude that bidding is futile, not participate, and avoid the exposure problem altogether. Of course, agents will not in general make perfect predictions. However, we find that even modestly informed predictions can significantly improve performance.

### 3.1 Bidding with price prediction

We now offer a straightforward way to modify SB with price prediction information, and in the following subsections present two easy-to-compute strategies that are instances of this approach.

Let $\mathbf{\Omega}$ be the set of information available to an agent that is relevant to predicting the prices of the $M$ goods. Partition $\mathbf{\Omega}$ as $(\mathbf{\Omega}_0; B)$, where B is the $t \times M$ history of bid prices revealed by the SAA as of the $t$th round, and $\mathbf{\Omega}_0$ is information available to the agent prior to the auction. Let $F \equiv F(\mathbf{\Omega}_0; B)$ denote a joint cumulative distribution function over final prices, representing the agent's belief. We assume that prices are bounded above by a known constant, $V$. Thus, $F$ associates cumulative probabilities with price vectors in $\{1, \ldots, V\}^M$.

We consider two ways to use prediction information to generate perceived prices differently from SB. In SB the agent calculates the best bundle evaluated at myopic perceived prices. We first define a *point predictor* for perceived prices, $\boldsymbol{\pi}$, which anticipates possible exposure risks. Then we define a *distribution predictor* for perceived prices, $\boldsymbol{\Delta}$, which in addition adjusts for the degree to which the agent's current winning bids are likely to be sunk costs.[3]

### 3.2 Point price prediction

Suppose the agent has (at least) point beliefs about the final prices that will be realized for each good. Let $\boldsymbol{\pi}(\mathbf{\Omega}_0; B)$ be a vector of predicted final prices. Before the auction begins the price predictors are $\boldsymbol{\pi}(\mathbf{\Omega}_0; \emptyset)$, where $\emptyset$ is the empty set of bid information available pre-auction.

The SAA reveals the current quotes each round. Since the auction is ascending, once the current bid price for good $m$ reaches $\beta_m$, there is zero probability that the final price $p_m$ will be less than $\beta_m$. We define a simple updating rule using this fact: the current price prediction for good $m$ is the maximum of the initial prediction and the myopically

---

[3]For both predictor strategies, if the agent has single-unit preference, it plays SB because that strategy is then no-regret.

perceived price:

$$\pi_m(\mathbf{\Omega}_0; B) \equiv \begin{cases} \max(\pi_m(\mathbf{\Omega}_0; \emptyset), \beta_m) & \text{if winning } m \\ \max(\pi_m(\mathbf{\Omega}_0; \emptyset), \beta_m + 1) & \text{otherwise.} \end{cases}$$

Armed with these predictions, the agent chooses the goods to bid on according to Equation 1 with $\hat{\boldsymbol{p}} \equiv \boldsymbol{\pi}(\mathbf{\Omega}_0; B)$. We denote a specific point price prediction strategy in this family by $PP(\boldsymbol{\pi}^x)$, where $x$ labels particular initial prediction vectors, $\boldsymbol{\pi}(\mathbf{\Omega}_0; \emptyset)$.

### 3.3 Distribution-based price prediction

Strategies using additional information from the distribution $F$ can at least weakly dominate strategies using only a predictor of the final price distribution mean. We assume the agent generates $F(\mathbf{\Omega}_0; \emptyset)$, an initial, pre-auction belief about the distribution of final prices.

As with the point predictor, we restrict the updating in our distribution predictor to conditioning the distribution only on the fact that prices must be bounded from below by $\boldsymbol{\beta}$. Let $\Pr(\boldsymbol{p} \mid B)$ be the probability, according to $F$, that the final price vector will be $\boldsymbol{p}$, conditioned on the information revealed by the auction, B. Then, with $\Pr(\boldsymbol{p} \mid \emptyset)$ as the pre-auction initial prediction, we define:

$$\Pr(\boldsymbol{p} \mid B) \equiv \begin{cases} \dfrac{\Pr(\boldsymbol{p} \mid \emptyset)}{\sum_{\boldsymbol{q} \geq \boldsymbol{\beta}} \Pr(\boldsymbol{q} \mid \emptyset)} & \text{if } \boldsymbol{p} \geq \boldsymbol{\beta} \\ 0 & \text{otherwise.} \end{cases} \quad (2)$$

(By $\boldsymbol{x} \geq \boldsymbol{y}$ we mean $x_i \geq y_i$ for all $i$.) For (2) to be well defined for all possible $\boldsymbol{\beta}$ we define the price upper bounds such that $\Pr(V, \ldots, V \mid \emptyset) > 0$.

We now use the distribution information to implement a further decision-theoretic modification to SB. If an agent is currently not winning a good and bids on it, then the expected incremental cost of winning the slot is the expected final price, with the expectation calculated with respect to the distribution $F$. If the agent is currently winning a good, however, then the expected incremental cost of winning that good depends on the likelihood that the current bid price will be increased by another agent, so that the first agent has to bid again to obtain the good. If, to the contrary, it keeps the good at the current bid, the full price is sunk (already committed) and thus should not affect incremental bidding. Based on this logic we define $\Delta_m$, the expected *incremental* price for good $m$.

First, for simplicity we use only the information contained in the vector of marginal distributions, $(F_1, \ldots, F_M)$, as if the final prices are independent across goods. Define the expected final price conditional on the most recent vector of bid prices, $\boldsymbol{\beta}$:

$$E_F(p_m \mid \boldsymbol{\beta}) = \sum_{q_m = \beta_m}^{V} \Pr(q_m \mid \beta_m) q_m.$$

The expected *incremental* price depends on whether the agent is currently winning good $m$. If not, then the lowest final price at which it could is $\beta_m + 1$, and the expected incremental price is simply the expected price conditional on $p_m \geq \beta_m + 1$,

$$\Delta_m^{\text{L}} \equiv E_F(p_m \mid \beta_m + 1)$$
$$= \sum_{q_m=\beta_m+1}^{V} \Pr(q_m \mid \beta_m + 1) q_m.$$

If the agent is winning good $m$, then the incremental price is zero if no one outbids the agent. With probability $1 - \Pr(\beta_m \mid \beta_m)$ the final price is higher than the current price, and the agent is outbid with a new bid price $\beta_m + 1$. Then, to obtain the good to complete a bundle, the agent will need to bid at least $\beta_m + 2$, and the expected incremental price is

$$\Delta_m^{\text{W}} = (1 - \Pr(\beta_m \mid \beta_m)) \sum_{q_m=\beta_m+2}^{V} \Pr(q_m \mid \beta_m + 2) q_m.$$

The vector $\mathbf{\Delta}$ of expected incremental prices is constructed by selecting $\Delta_m^{\text{L}}$ or $\Delta_m^{\text{W}}$ respectively for each $m$, depending on whether the agent's bid is currently winning. To select the bundle on which it will bid, the agent evaluates Equation 1 with $\hat{p} \equiv \mathbf{\Delta}$. We denote the strategy of bidding based on a particular distribution predictor by $PP(F^x)$, where $x$ labels various distribution predictors, $F(\mathbf{\Omega}_0; \emptyset)$.

## 4 Self-Confirming Price Distributions

### 4.1 Definition and existence

An *SAA environment* comprises an SAA mechanism over $M$ goods, a set of $N$ agents, and a probability distribution over $M$-good value functions for each agent.

**Definition 1** *Let SE be an SAA environment. The prediction F is a self-confirming price distribution for SE iff F is the distribution of prices resulting when all agents play bidding strategy $PP(F)$ in SE.*

A prediction is *approximately self-confirming* if the definition above is satisfied for some approximate sense of equivalence between the outcome and prediction distributions.

The key feature of self-confirming prices, of course, is that agents make decisions based on predictions that turn out to be correct with respect to the underlying probability distribution. Since agents are optimizing for these predictions, we might reasonably expect the strategy to perform well in an environment where its predictions are confirmed.

The actual joint distribution will in general have dependencies across prices for different goods. We are also interested in the situation in which if the agents play a strategy based just on marginal distributions, that resulting distribution has the same marginals, despite dependencies.

**Definition 2** *Let SE be an SAA environment. The prediction $F = (F_1, \ldots, F_M)$ is a vector of self-confirming marginal price distributions for SE iff for all m, $F_m$ is the marginal distribution of prices for good m resulting when all agents play bidding strategy $PP(F)$ in SE.*

Note that the confirmation of marginal price distributions is based on agents using these predictions as if the prices of goods were independent. However, we consider these predictions confirmed in the marginal sense as long as the results agree for each good separately, even if the joint outcomes do not validate the independence assumption.

A natural question to raise at this point is whether self-confirming predictions can actually be identified in plausible SAA environments. We demonstrate below that we can often find approximately self-confirming marginal predictions. However, it is easy to show that they cannot generally exist, by invoking a particular case known to be difficult for SAAs. Specifically, consider the $M = N = 2$ configuration presented in Table 4.1, a common illustration of the absence of a competitive equilibrium (Cramton, 2005). There exist no prices for goods 1 and 2 such that both agents optimize their demands at the specified prices and the markets clear.

Table 2: A configuration with no price equilibrium.

| Name | $v(\{1\})$ | $v(\{2\})$ | $v(\{1,2\})$ |
|---|---|---|---|
| Agent 1 | 0 | 0 | 30 |
| Agent 2 | 20 | 20 | 20 |

**Proposition 1** *There exist SAA environments for which no self-confirming or marginally self-confirming price distributions exist.*

*Proof.* Define an SAA corresponding to the configuration of Table 4.1. Given a deterministic SAA mechanism (one without asynchrony or random tie-breaking), for fixed value functions the outcome from playing any profile of deterministic trading strategies is a constant. Thus, the only possible self-confirming distributions (which were defined for agents playing the deterministic $PP(F)$ strategies) must assign probability one to the actual resulting prices. But given such a prediction, our trading strategy will pursue the agent's best bundle at those prices, and must actually get them since the prices are correct if the distribution is indeed self-confirming. But then the markets would all clear, contrary to the fact that the predicted prices cannot constitute an equilibrium, since such prices do not exist in this instance. □

Despite this negative finding, we conjecture that approximately self-confirming price distributions exist for a large class of nondegenerate preference distributions, and can be computed given a specification of the preference distribu-tion. We now present a procedure for deriving such distributions, and some evidence for its effectiveness.

### 4.2 Deriving self-confirming price distributions

Given an SAA environment—including the distributions over agent preferences— we derive self-confirming price distributions through an iterative simulation process. Starting from an arbitrary prediction $F^0$, we run many instances of an SAA environment (sampling from the given preference distributions) with all agents playing strategy $PP(F^0)$.[4] We record the resulting prices from each instance, and designate the sample distribution observed by $F^1$. We then run another battery of instances, or *iteration*, with agents playing $PP(F^1)$, and repeat the process in this manner for some further series of iterations. If it ever reaches an approximate fixed point, with $F^t \approx F^{t+1}$ for some $t$, then we have statistically identified an approximate self-confirming price distribution for this environment. (Due to sampling error, the approximate version of the concept is the best we can attain through simulation.)

Any reasonable measure of similarity of probability distributions combined with a threshold constitutes an operable policy for validating approximate self-confirmation. We employ the Kolmogorov-Smirnov (KS) statistic, defined as the maximal distance between any two corresponding points in the CDFs:

$$KS(F,F') = \max_x |F(x) - F'(x)|.$$

When we seek self-confirmation only of predictors for the marginal distributions, we measure KS distance separately for each good, and take the largest value: $KS_{\text{marg}} \equiv \max_m KS(F_m, F'_m)$.

Our complete procedure for deriving approximate self-confirming price distributions is defined by specifying:

1. a number of samples per iteration,

2. a threshold on $KS$ or $KS_{\text{marg}}$ on which to halt the iterations and return a result,

3. a maximum number of iterations in case the threshold is not met,

4. a smoothing parameter $k$ designating a number of iterations to average over when the procedure reaches the maximum iterations without finding an approximate fixed point.

The bound on the number of iterations ensures that this procedure terminates and returns a price distribution, which may or may not be self-confirming. When this occurs, the smoothing parameter avoids returning a distribution that is known to cause oscillation. However, the apparent nonexistence of a self-confirming equilibrium in this case suggests the problem cannot be totally eradicated, and we do not expect the strategy to perform as well when the underlying oscillations are large.

---
[4]In our experiments the initial prediction is zero prices, but our results do not appear sensitive to this.

Table 3: Descriptive Statistics for a Self-confirming Price Distribution Calculated in Six Iterations

| Good | Mean Price | Standard Deviation |
|---|---|---|
| 1 | 10.8 | 7.7 |
| 2 | 6.5 | 5.1 |
| 3 | 4.1 | 3.7 |
| 4 | 2.3 | 2.5 |
| 5 | 1.0 | 1.3 |

### 4.3 Trials of the iterative procedure

To illustrate the process, we specify an SAA scheduling environment with five agents competing for five time slots. An agent's value function is defined by its job length and its value for meeting various deadlines. We draw job lengths randomly from $U[1,5]$. We choose deadline values randomly from $U[1,50]$ then prune to impose monotonicity; for details see Reeves et al. (2005). We set the algorithm parameters at one million games per iteration, and a 0.01 KS convergence criterion.[5] We ran the algorithm, playing the million games per iteration to generate an empirical price distribution. The predicted and empirical distributions quickly converged, with a KS statistic of 0.007 after only six iterations. We report descriptive statistics for the result in Table 3.

To see if our method produces useful results with some regularity, we applied it to 22 additional instances of the scheduling problem, varying the numbers of agents and goods, and the preference distributions. We again drew deadline values from $U[1,50]$ and pruned them for monotonicity. We used two probability models for job lengths in the first 21 instances. In the *uniform* model, they are drawn from $U[1,M]$. In the *exponential* model job length $\lambda$ has probability $2^{-\lambda}$, for $\lambda = 1, \ldots, M-1$, and probability $2^{-(M-1)}$ when $\lambda = M$.

We constructed 10 instances of the uniform model, comprising various combination of $3 \leq N \leq 9$ and $3 \leq M \leq 7$. In each case, our procedure found self-confirming marginal price distributions (KS threshold 0.01) within 11 iterations. Similarly, for 11 instances of the exponential model, with the number of agents and goods varying over the same range, we found SC distributions within seven iterations. We plot the distribution of KS values from these 21 in-stances in Figure 1.

---
[5]Since KS is a distance between CDFs, a 0.01 threshold is equivalent to a maximal one percentage point probability difference at any point in the two distributions.

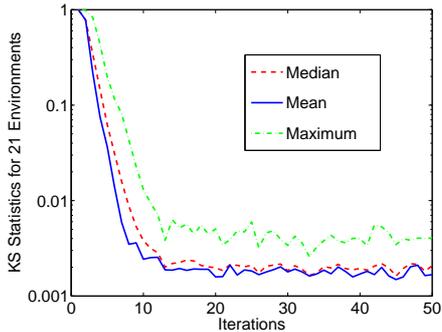

Figure 1: Convergence of iterative SC price estimation.

The 22nd instance was designed to be more challenging: we used the $N = M = 2$ example with fixed preferences described in Table 4.1. Since there exists no SC distribution, our algorithm did not find one, and as expected after a small number of iterations it began to oscillate among a few states indefinitely. After reaching the limit of 100 iterations, our algorithm returned as its prediction distribution the average over the last $k = 10$.

## 5 Empirical Game Analysis

We now analyze the performance of self-confirming distribution predictors in a variety of SAA environments, against a variety of other strategies. We use methods developed in our prior work (Reeves et al., 2005; MacKie-Mason et al., 2004), and related to recent studies in a similar empirical vein (Armantier et al., 2000; Kephart et al., 1998; Walsh et al., 2002).

### 5.1 Environments and strategy space

We studied SAAs applied to market-based scheduling problems, as described in Section 4.2. Particular environments are defined by specifying the number $M$ of goods, the number $N$ of agents, and a preference model comprising probability distributions over job lengths and deadline values. The bulk of our computational effort went into an extensive analysis of one particular environment, the $N = M = 5$ uniform model presented above. As described in Section 5.2, the empirical game for this setting provides much evidence supporting the unique strategic stability of $PP(F^{SC})$. We complement this most detailed trial with smaller empirical games for a range of other scheduling-based SAA environments. Altogether, we studied selected environments with uniform, exponential, and fixed distributions for job lengths; a modified uniform distribution for deadline values; and agents in $3 \leq N \leq 8$; goods in $3 \leq M \leq 7$.

To varying degrees, we analyzed the interacting performance of 53 different strategies. These were drawn from the three strategy families described above: SB, point predictor, and distribution predictor. For each family we varied a defining parameter to generate the different specific strategies.[6] The choice of strategies was based on prior experience, though we make no claim that we covered all reasonable variations. Naturally, our emphasis is on evaluating the performance of $PP(F^{SC})$ in combination with the other strategies. One of the noteworthy alternatives is $PP(F^{SB})$, which employs the price distribution prediction formed by estimating (through simulation) the prices resulting from all agents playing SB.

We use Monte Carlo simulation to estimate the *payoff function* for our empirical game, which maps profiles of agent strategies to expected payoffs for each agent. Given $N$ agents and $S$ possible strategies, the symmetric game comprises $\binom{N+S-1}{N}$ distinct strategy profiles. In our primary example with five agents, there are over four million different strategy profiles to evaluate. Since we determine the expected payoffs empirically for each profile by running millions of simulations of the auction protocol, estimating the entire payoff function is infeasible. However, we can estimate the complete payoff matrix for various subsets of our 53 strategies. As we describe below, we do not need the full payoff matrix to reach conclusions about equilibria in the 53-strategy game.

### 5.2 5×5 uniform environment

The largest empirical SAA game we have constructed is for the SAA scheduling environment discussed in Section 4.2, with five agents, five goods, and uniform distributions over job lengths and deadline values. We have estimated payoffs for 4916 strategy profiles, out of the 4.2 million distinct combinations of 53 strategies. Payoff estimates are based on an average of 10 million samples per profile (though some profiles were simulated for as few as 200 thousand games, and some for as many as 200 million). Despite the sparseness of the estimated payoff function (covering only 0.1% of possible profiles), we obtained several results.

First, as discussed above, we conjectured that the self-confirming distribution predictor strategy, $PP(F^{SC})$, would perform well. We directly verified this: *the profile where all five agents play a pure $PP(F^{SC})$ strategy is a Nash equilibrium* in the game restricted to 53 strategies. No unilateral deviation to any of the other 52 pure strategies is profitable. To verify a pure-strategy symmetric equilibrium (all agents playing $s$) for $N$ players and $S$ strategies, one needs payoffs for only $S$ profiles: one for each strategy playing with $N-1$ copies of $s$. The symmetric profile is an equilibrium if there are no profitable deviation profiles (i.e., obtained by changing the strategy of one player to obtain a higher payoff given the others).

---

[6]To conserve space, we have an appendix available upon request with a full description of the 53 strategies here. An appendix is in preparation, to provide specification of all the parameters, including description of the prediction methods used for point and distribution predictors.

The fact that $PP(F^{SC})$ is pure symmetric Nash for this game does not of course rule out the existence of other Nash equilibria. Indeed, without evaluating any particular profile, we cannot eliminate the possibility that it represents a (non-symmetric) pure-strategy equilibrium itself. However, the profiles we did estimate provide significant additional evidence, including the elimination of broad classes of potential symmetric mixed equilibria.

Let us define a strategy *clique* as a set of strategies for which we estimated payoffs for all combinations. Each clique defines a subgame, for which we have complete payoff information. Within our 4916 profiles we have eight maximal cliques, all of which include strategy $PP(F^{SC})$. For each of these subgames, $PP(F^{SC})$ is the only strategy that survives iterated elimination of (strictly) dominated (pure) strategies. It follows that $PP(F^{SC})$ is the unique (pure or mixed strategy) Nash equilibrium in each of these clique games. We can further conclude that in the full 53-strategy game there are no mixed strategy equilibria with support contained within any of the cliques, other than the special case of the pure-strategy $PP(F^{SC})$ equilibrium.

We can show that $PP(F^{SC})$ is the only small-support mixed strategy, among the two-strategy cliques for which we have calculated payoffs, that is even an approximate equilibrium. Of the $\binom{52}{2}=1326$ pairs of strategies not including $PP(F^{SC})$, we have all profile combinations for 46. For these we obtained a lower bound of 0.32 on the value of $\varepsilon$ such that a mixture of one of these pairs constitutes an $\varepsilon$-Nash equilibrium. In other words, for any symmetric profile defined by such a mixture, an agent can improve its payoff by a minimum of 0.32 through deviating to some other pure strategy. For reference, the payoff for the all-$PP(F^{SC})$ profile is 4.51, so this represents a nontrivial difference.

Finally, for each of the 4916 evaluated profiles, we can derive a bound on the $\varepsilon$ rendering the profile itself an $\varepsilon$-Nash pure-strategy equilibrium. The three most strategically stable profiles by this measure (i.e., lowest potential gain from deviation, $\varepsilon$) are:

1. all $PP(F^{SC})$: $\varepsilon = 0$ (confirmed Nash equilibrium)

2. one $PP(F^{SB})$, four $PP(F^{SC})$: $\varepsilon > 0.13$

3. two $PP(F^{SB})$, three $PP(F^{SC})$: $\varepsilon > 0.19$

The remaining profiles have $\varepsilon > 0.25$.

Our conclusion from these observations is that $PP(F^{SC})$ is a highly stable strategy within this strategic environment, and likely uniquely so. Of course, only limited inference can be drawn from even an extensive analyis of only one particular distribution of preferences.

### 5.3 Self-confirming prediction in other environments

To test whether the strong performance of $PP(F^{SC})$ generalizes across other SAA environments, we undertook smaller versions of this analysis on variations of the model above. Specifically, we explored 17 additional instances of the market-based scheduling problem: eight each with the uniform (U) and exponential (E) models (3–8 agents, 3–7 goods), and one with fixed preferences, corresponding to the counterexample model of Table 4.1.

For each we derived self-confirming price distributions (failing in the last case, of course), as reported in Section 4.3. We also derived price vectors and distributions for the other prediction-based strategies. For 11 of the environments (eight U and three E), we evaluated 27 profiles: one with all $PP(F^{SC})$, and for each of 26 other strategies $s$, one with $N-1$ agents playing $PP(F^{SC})$ and one agent playing $s$. We ran between two and ten million games per profile in all of these environments.

For each of these 11 U and E models, we identified the seven best responses to others playing $PP(F^{SC})$ (which invariably included $PP(F^{SC})$ itself). To economize on simulation time, for the other five E environments we used the seven best responses found for the most similar of the simulated E environments. We then evaluated all profiles involving these strategies (i.e., a 7-clique) in the for each of the 16 environments, based on at least 340,000 samples per profile.

We summarize our results for U and E models in Table 4. We first report the percentage gain for a participant that deviates from all-$PP(F^{SC})$ to the best of the 26 other strategies. (This is the value of $\varepsilon$ from an $\varepsilon$-Nash equilibrium as a percent of the average payoff from all-$PP(F^{SC})$ play.) To adjust for the sampling error in our method, we report in the second column the average percentage gain from deviation when we bootstrap a sample of 30 observations from a normal distribution with the payoff mean and variance we observed in our simulations. Using the same sampling distribution, we calculate the probability that all-$PP(F^{SC})$ is a Nash equilibrium. Finally, for each environment with $N \leq 6$ we used replicator dynamics to find a mixed strategy equilibrium for the 7-clique, and report the probability that an agent plays $PP(F^{SC})$ in this Nash equilibrium.

In 15 out of these 16 environments, $PP(F^{SC})$ was verified to be an $\varepsilon$-Nash equilibrium for an $\varepsilon$ less than 2% of the average payoff. That is, no single agent could gain as much as 2% by deviating. The worst performance by $PP(F^{SC})$ is in environment $U(7,8)$, for which a strategy deviation could improve expected payoff by only 5%. Further, our results are quite insensitive to sampling error induced by our simu-lation method. The sampling-adjusted percentage gain values are never significantly higher, and in every environment for which we have a mixed-strategy equilibrium, $PP(F^{SC})$ appears with substantial if not overwhelming probability.

Table 4: Performance of All-$PP(F^{SC})$ as a Candidate Equilibrium for Various U and E Environments

| Env($M,N$) | Percentage gain from one-player deviation | Percentage gain adjusted for sampling error | Probability all-$PP(F^{SC})$ is exact Nash equilibrium | Mixed-strategy probability of playing $PP(F^{SC})$ |
|---|---|---|---|---|
| $E(3,3)$ | 0 | 0 | 1.00 | 1.00 |
| $E(3,5)$ | 0 | .09 | .600 | .996 |
| $E(3,8)$ | .83 | .85 | 0 | — |
| $E(5,3)$ | 0 | 0 | 1.00 | .999 |
| $E(5,5)$ | 0 | .01 | .900 | .998 |
| $E(5,8)$ | .60 | .64 | 0 | — |
| $E(7,3)$ | 0 | .06 | .667 | .992 |
| $E(7,6)$ | .04 | .10 | .567 | .549 |
| $U(3,3)$ | 1.24 | 1.26 | 0 | .725 |
| $U(3,5)$ | 0 | 0 | 1.00 | 1.00 |
| $U(3,8)$ | .56 | .53 | 0 | — |
| $U(5,3)$ | 1.35 | 1.35 | 0 | .809 |
| $U(5,8)$ | 1.59 | 1.62 | 0 | — |
| $U(7,3)$ | .81 | .84 | 0 | .942 |
| $U(7,6)$ | .52 | .52 | 0 | .929 |
| $U(7,8)$ | 4.98 | 4.94 | 0 | — |

Overall, we regard this as favorable evidence for the $PP(F^{SC})$ strategy across the range of market-based scheduling environments. Not surprisingly, the environment with fixed preferences is an entirely different story. Recall that in this case the iterative procedure failed to find a self-confirming price distribution, and the averaged distribution computed by our algorithm provides a quite inaccurate prediction. For this environment we evaluated all 53 profiles with at least one agent playing $PP(F^{SC})$. The self-confirming prediction strategy performed poorly, generally obtaining negative payoffs regardless of other strategies. Since one available strategy is to simply not trade, $PP(F^{SC})$ is clearly not a best-response player in this environment.

## 6 Discussion

Our proposed trading strategy for SAA environments with complementarities places bids based on probabilistic predictions of final good prices. Like the approach of Greenwald and Boyan (2004), our policy tackles the exposure problem head-on, by explicitly weighing the risks and benefits of placing bids on alternative bundles, or no bundle at all. The specific strategy generalizes our previous work on bidding based on point price predictions, and like that scheme is parametrized by the *method* for generating predictions. By explicitly conditioning on context (e.g., preference distributions), such trading strategies are potentially robust across varieties of SAA environments.

The strategy we consider most promising employs what we call *self-confirming price distributions*. A price distribution is self-confirming if it reflects the prices generated when all agents play the trading strategy based on this distribution. Although such self-confirming distributions may not always exist, we expect they will (at least approximately) in many environments of interest, especially those characterized by relatively diffuse uncertainty and a moderate number of agents. An iterative simulation algorithm appears effective for deriving such distributions.

Given the analytic and computational intractability of the game induced by an SAA environment, we evaluated our approach using an empirical game-theoretic methodology. We explored a restricted strategy space including a range of candidate strategies identified in prior work. Despite the infeasibility of exhaustively exploring the profile spaces, our analyses support several game-theoretic conclusions. The results provide favorable evidence for our favored strategy—very strong evidence in one environment we investigated intensely, and somewhat less categorical evidence for a range of variant environments.

Neither the strategy we present here nor any other strategy is likely to be universally best across SAA environments. Nevertheless, our results establish the self-confirming price prediction strategy as the leading contender for dealing broadly with the exposure problem. If agents make optimal decisions with respect to prices that turn out to be right, there may not be room for performing a lot better. On the other hand, there are certainly areas where improvement

should be possible, for example:

- accounting for one's own effect on prices, as in strategic demand reduction (Weber, 1997)
- incorporating price dependencies with reasonable computational effort
- more graceful handling of instances without self-confirming price distributions
- timing of bids: trading off the risk of premature quiescence with the cost of pushing prices up

We intend to explore some of these opportunities in our continuing research.

# Acknowledgments

Rahul Suri contributed to the equilibrium analysis in Section 5. The anonymous reviewers suggested helpful clarifications. This work was supported in part by National Science Foundation grant IIS-0414710.